# T U M

## I N S T I T U T   F Ü R   I N F O R M A T I K

Semantics of UML

## Towards a System Model for UML

The Structural Data Model

Version 1.0

**Manfred Broy, María Victoria Cengarle, Bernhard Rumpe**

with special thanks to

**Michelle Crane, Jürgen Dingel, Zinovy Diskin, Jan Jürjens, Bran Selic**

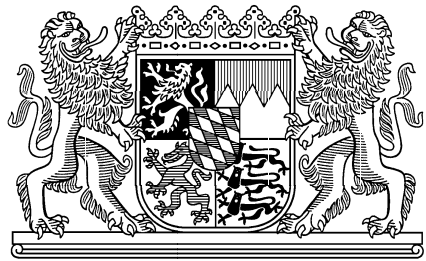

TUM-I0612
Juni 06

## T E C H N I S C H E   U N I V E R S I T Ä T   M Ü N C H E N

Semantics of UML

# Towards a System Model for UML

## The Structural Data Model

Version 1.0

**Manfred Broy[1]**
**María Victoria Cengarle[1]**
**Bernhard Rumpe[2]**

[1] Lehrstuhl für Software & Systems Engineering
Institut für Informatik
Technische Universität München

[2] Institut für Software Systems Engineering
Carl-Friedrich-Gauß-Fakultät für Mathematik und Informatik
Technische Universität Braunschweig

**with special thanks to**
**Michelle Crane**
**Jürgen Dingel**
**Zinovy Diskin**
**Jan Jürjens**
**Bran Selic**

# Table of Contents

# 1   Introduction into the System Model for UML

In this document we introduce a system model as the basis for a semantic model for UML 2.0. The system model is supposed to form the core and foundation of the UML semantics definition. For that purpose the basic system is targeted towards UML.

This document is structured as follows: In the rest of Section 1 we will discuss the general approach and highlight the main decisions. This section is important to understand the rest of this document. Section 2 contains the actual definition of the structural part of the system model. It is built in layers as described in Section 1. For brevity of the approach, we defer deeper discussions into the Appendix in Section 4.

This document is part of a project on the formalization of the UML 2.0 in cooperation between the Queens University Kingston and the Technische Universitäten Braunschweig and München. This version 1.0 is the result of a longer effort to define the structure, behaviour and interaction of object-oriented, possibly distributed systems abstract enough to be of general value, but also in sufficient detail for a semantic foundation of the UML. We also wish to thank external reviewers, and especially Gregor von Bochmann, Gregor Engels and Sébastien Gérard for their help.

## 1.1   General Approach to Semantics

The semantics of any formal language consists of the following basic parts [Win93]:

- the syntax of the language in question (here: UML) – be it graphical or textual,

- the semantic domain, a domain well known and understood based on a well-defined mathematic theory, and

- the semantic mapping: a functional or relational definition that relates both, the elements of the syntax and the elements of the semantic domain.

This technique of giving meaning to a language is the basic principle of semantics: every syntactic construct is mapped onto a semantic construct. As discussed in the literature, there are many flavours of these three elements.

Normally, the set of syntactic elements and the semantic domain bear some structure, and the semantic mapping desirably preserves or is compatible with this structure. Mathematically, this aspect is crucial and is denominated compositionality.

In [BKR96] the term system model was used the first time to describe a semantic domain; it defines a family of systems, describing their structural and behavioural issues. Each concrete syntactic instance (in our case, an individual UML diagram, or even a part of it) is interpreted by the semantic mapping as a predicate over the set of systems defined by the system model.

As explained in [HR04] the semantic mapping does have a form:

Sem: UML $\rightarrow$ $\wp$ (Systemmodel)

and thus functionally relates any item in the syntactic domain to a construct of the semantic domain. The semantics of a model m ∈ UML is therefore Sem(m). [4.1][1]

The system model described in this report identifies the set of all possible OO systems that can be defined using a subset of UML which we call "clean UML" as introduced below.

## 1.2  Structuring the Semantics of UML

The overall goal of giving semantics to a graphical modelling language is depicted in Fig. 1. The basic idea expressed by this diagram is as follows: The full graphical language is related to a simplified language by some transformations that allow us to get rid of some notational extensions and derived concepts by reducing them to constructs of the simplified language.

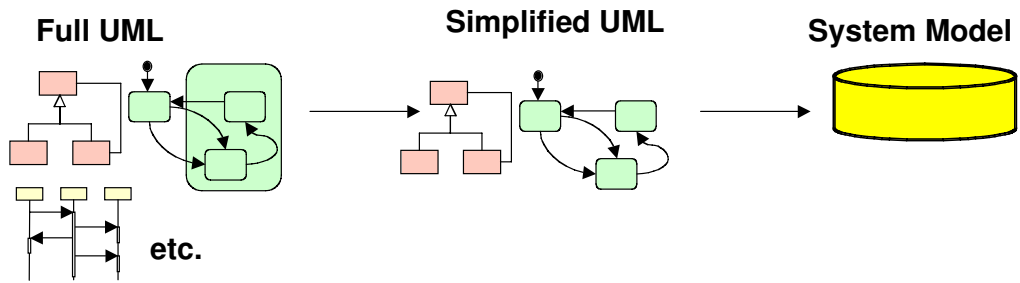

**Full UML**          **Simplified UML**          **System Model**

etc.

**Fig. 1** General strategy for the definition of the semantics of UML 2.0

Actually, the semantic mapping may possibly not deal with full UML in the end but with a cleaned subset of UML.  Concretely, to deal with the complexity of defining a proper semantics, we opt for a decomposition of the mapping

Sem: UML → ℘(Systemmodel)

in several ways:

- Full UML is restricted to a subset (called "clean UML") that can be treated semantically without overly sophisticated constructs.

- Clean UML is mapped by transformations into a simplified UML core (see Fig. 1, left side). In doing so, derived constructs of UML are replaced by their definition in terms of constructs of the core. [4.2]

- Simplified UML, finally, is mapped to the system model using a predicative approach (see Fig. 1, right side).

As mentioned above, the system model describes the "universe (set) of all possible semantic structures (each with its behaviour)". The semantic mapping interprets a UML model as a predicate that restricts the universe to a certain set of structures, which represents the meaning of the UML model.

---

[1] Within square brackets we refer the interested reader to deeper comments in the appendix.

In other words, our goal is to define the semantics of a comprehensive core of well-defined concepts of UML, and not of all its notational extensions. This way of proceeding is called the onion approach. The system model, i.e., the right hand side of the semantics mapping Sem, does not address UML and its constructs, it is targeted towards the left hand side of Sem, namely (simplified) UML 2.0, by covering a number of basic concepts expressible in the UML. [4.2]

The system model defines a universe of interacting state machines that describe the behaviour of objects and embody their data structure. This universe is introduced in layers as shown in Fig. 2.

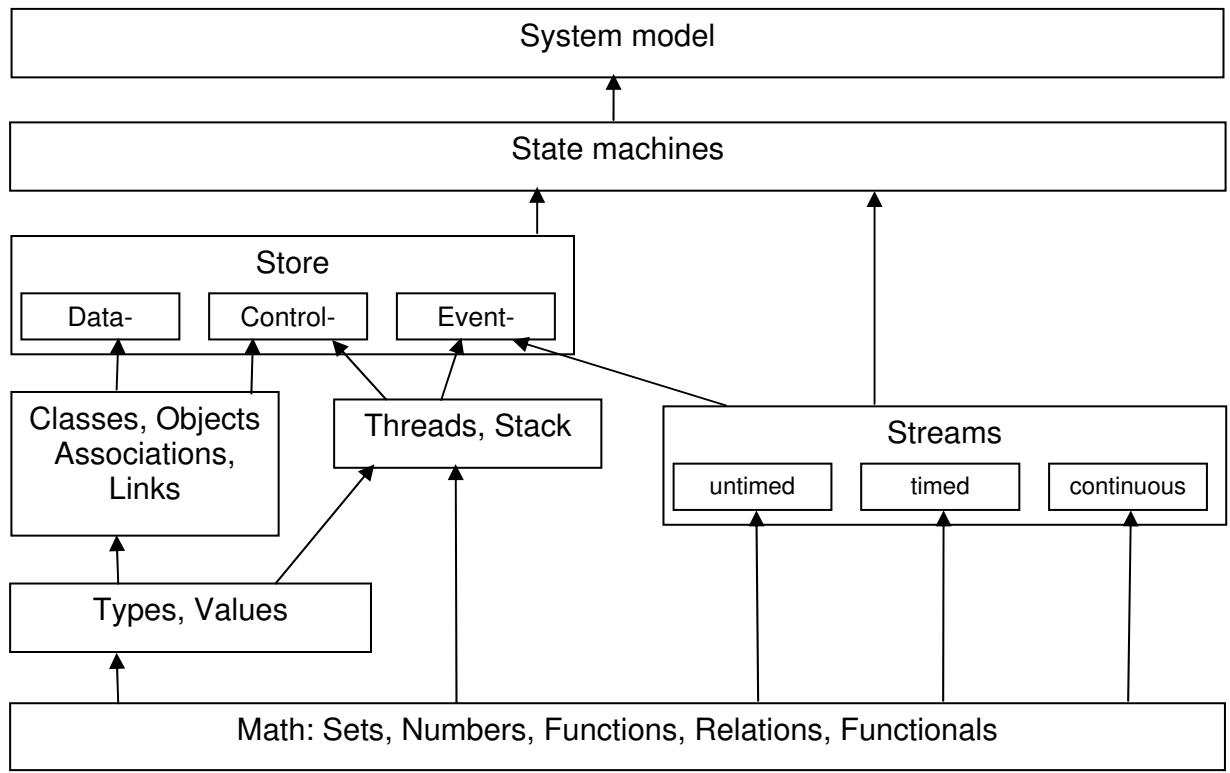

**Fig. 2** Theories that constitute the system model

This first report focuses on the structural characteristics of the system. Other parts of the system model cover the state transition and the control parts; these will be introduced in separate reports.

When describing the concepts, we will introduce below, precisely, some decisions have to be made that can be left open or do not even occur when staying informal. We clearly identify those decisions either directly, or mark them as a "variation point" or as a possible "extension point" and leave it to the user of the system model to adopt a variation or extension.

The rectangles in Fig. 2 contain mathematical concepts, whereas arrows show a relationship among concepts that could be paraphrased as "is defined in terms of". For instance, type names (not to be confused with the syntactic entity of types of a programming language) and values are used to define, on the one hand, classes and objects and, on the other, threads and stacks.

## 1.3 The Math behind the System Model

A precise description of the system model calls for a precise instrument. For our purposes, mathematics is exactly appropriate because of its power and flexibility. Admittedly, sometimes mathematics is not quite intuitive and thus needs readers to cope with it. Using UML itself to describe semantics of UML might seem, on the contrary, a pragmatic approach. This approach is somewhat meta-circular and necessarily calls for a kind of bootstrap, typically mathematics again. Moreover, understanding the semantics of UML in terms of UML itself, demands a very good knowledge of the language whose semantics is about to be formally given. Besides, UML does not conveniently provide the appropriate mechanisms we need. Because of these reasons we decided to use only mathematics. Of course, whenever appropriate, we use diagrams to illustrate some mathematically defined concepts, but the diagrams do not replace the mathematical formulas. [4.3]

The following principles have proven to be useful when defining the system model:

1. Pure math is used to define the system model. Its sub-theories are built on: numbers, sets, relations, and functions. Additional theories are built in layered form; see Fig. 2. That is, only notation and mathematical definitions and neither new syntax nor language are introduced or used in the system model. Diagrams are occasionally used to clarify things, but do not formally contribute to the system model.

2. The system model does not constructively define its elements, but introduces the elements and characterises their properties. That is, abstract terms are used whenever possible. For instance, instead of using a record to define the structure of an object, we can use an abstract set and a number of selector functions. Properties of the set are then defined through such selectors. This however might pose a problem to "constructivists" who want everything constructively developed. Based on our background and knowledge, we claim that we could transform this system model into a constructive version, but that would be more awkward to read and less intuitive, as it costs a lot more machinery, like fixed points, etc.

3. Everything important is given an appropriate name. For instance, in order to deal with classes, there is a "universe of class names" UCLASS, and similarly there is also a "universe of type names" UTYPE, which however is just a set of *names* (and not types); see Sects. 2.6 and 2.1 below.

4. To our best knowledge any underlying assumptions were avoided, according to the slogan: What is not explicitly specified needs not hold. If we for instance do not explicitly state that two sets are disjoint, these two sets might have elements in common: e.g., we do not enforce the set of type names UTYPE and the set of values UVAL to be disjoint (but also do not take advantage of a possible overlap); see Sect. 2.1 below. Sometimes these loose (underspecified) ends are helpful to specialise the system model and are there on purpose.[2] If you need a property, (a) check whether it is there, (b) if absent, check whether it can be inferred as emerging property, (c) if not, do you really need it?, and (d), if yes, you may add it as an additional restriction.

---

[2] Note that this would not have been possible when defining the system model in a constructive way.

5. Generally, deep embedding (or explicit representation) is used. This means the semantics of the embedded language, i.e., UML, is completely formalised within the supporting language, in our case, mathematics.[3] As a consequence, the system model does not have and does not need a type system itself. However, it characterises the type system of UML.

6. Dynamic sets, like the set of actually existing object identifiers in a certain snapshot of a system execution, are modelled in two parts: A universe of object identifiers describes the maximal set (which is infinite), and a finite set of existing object identifiers is part of the system state.

7. Specific points, where the system model could be further strengthened, have been marked as "extension" and "variation points". Extensions deal with additional elements that can be defined upon the system model. Extension points allow us to show additional machinery that need not present in each modelled system. Prominent examples of such extensions are the existence of a predefined top-level class called "Object" or an enhanced type system, including e.g., templates.

8. Variations describe changes of definitions, that lead to a slightly different system model . Variation points allow us to describe specialized variants of the system model, that may not be generally valid, but hold for a large part of possible systems. Prominent examples are single inheritance hierarchies or type-safe overriding of operations in subclasses, which are not given in UML in general.

## 1.4  Static and Dynamic Issues

An object-oriented system can basically be described using one of various existing paradigms. We opted for the paradigm of a global state machine[4] in order to accommodate a global (and maybe distributed) state space. A UML model will be interpreted thus as a single state machine encompassing, among other things, the meaning of the state transition diagrams within the model. The system model, thus, defines a universe of state machines. A state machine is given by its state space, its initial states, and its state transition function. [4.4] Please note that our notion of state machine is more basic and does not directly relate to the state machines/state transition diagrams the UML provides.

The types and classes of a UML model are static, i.e., they do not change over the lifetime of a system. This information is called the static information of a state machine. The set of existing objects and the values of the attributes as well as some of the relations are dynamic, i.e., they may change in transition steps. This latter is called the dynamic information of a state machine and is coded in the states of the state machine. The set of invoked and not (fully) executed methods is administrated in the control part of a state machine.

---

[3] Shallow embedding, on the contrary, means that the mapping from the interpreted to the interpreting language is defined in some suitable meta-language.

[4] The term "state machine" refers to the members of the universe defined by the system model. This term is conceptually similar to UML's state transition diagrams, but not formally related. One may also think of an "abstract implementation" when referring to such a state machine.

Summarising, a state machine of the system model is constituted of the following elements:

- Static part: class names and type names definitions [4.4]

- Dynamic part: set of created objects and the state of their attributes [4.4]

- Control part: invoked methods [4.4]

- State transition function: defined in terms of the control part (not treated in this report) [4.4]

- Interface abstraction: encapsulation of the local state.

Changes to the static part, that arise when e.g., the UML model evolves or is reconfigured,[5] are not considered in this report.

In the next section we detail the static part of any global state machine of the system model.

## 1.5  What is the system model?

A system model provides a means to define the semantics of any UML model. In precise mathematical terms, the states of a state machine of the system model are defined in terms of a larger number of mathematically defined elements that are subsequently introduced.

---

**Definition:** The system model SYSMOD is the universe of state machines whose

- states are triples (DataStore, ControlStore, EventStore), where
  - the DataStore, defined below, depends on the static structure, and
  - the ControlStore and the EventStore are defined in separate documents;
- state transition function is also defined in a separate document.

---

Formally, when speaking about a state machine of the system model, we speak of an instance sm∈SYSMOD. More precisely, a universe UTYPE of type names (as will be introduced in Sect. 2 below) defined for sm∈SYSMOD, is not necessarily unique; it is shorthand for sm.UTYPE meaning that UTYPE is the universe of type names of the state machine sm. We simply abbreviate to UTYPE whenever sm is clear from the context.

---

[5] The changes meant here are closely related to the schema evolution of databases.

## 1.6 What will be next for the system model?

This document contains the structural part of the system model. The next two parts, each in a separate document, will deal with control (processes, communication, etc.) and with a state-based/interaction definition for the system model.

Parallel to having designed the system model, we are now going into the process of using it to define semantics for the most important notations of the UML. Along with this process of defining UML semantics, we hope to be able to enhance the system model defined in these three parts further, to lay a solid and also generally acceptable semantic basis for the UML.

## 2 Static Part of the System Model

In this section, we introduce the fundamental static part of the state machines in the system model that will serve to define the semantics of UML models.

Formally, the static part of a state machine in the system model is composed of, among other things: [4.5]

- UTYPE, the universe of type names,

- UVAL, the universe of values,

- UCLASS, the universe of class names, and

- UOID, the universe of object identifiers.

Recall Fig. 2, whose elements are defined and put in relationships below. Please note that we do not further prescribe what "names" are but take them as primitives.

## 2.1 Type Names and Their Carrier Sets

A type name identifies a carrier set, which contains simple or complex data elements called members or values of (or associated with) the type name. Members of all type names are gathered in the universe UVAL of values. Formally, [4.5]

---

**Definition**

- UTYPE is the universe of type names

- UVAL is the universe of values

- CAR: $UTYPE \rightarrow \wp(UVAL)$ maps type names to associated carrier sets

- For $T \in UTYPE$ and $e \in CAR(T)$ the pair $(T, e)$ denotes a typed element of type name $T$

---

*Variation Point:* In a very general fashion, we do not enforce carrier sets to be disjoint or values to know to which carrier set they belong. For certain type names we may even assume that their carrier sets are identical or in a subset relation. [4.5]

*Remark:* As an aside, note that, in a proper typing system, families of types, together with their functions, form algebras with specific signatures. For details see the concept of abstract data types [Bro98].

## 2.2   Basic Type Names and Type Name Constructors

We assume a number of basic type names for basic values are given. We only require type names for Boolean and integer values.

---

**Definition**

- Bool, Int $\in$ UTYPE,

- CAR(Boolean) = {true,false} with true, false $\in$ UVAL and true $\neq$ false

- CAR(Int) = $\mathbb{Z} \subseteq$ UVAL is the set of integer values

- $\approx \subseteq$ UTYPE $\times$ UTYPE is an equivalence relation on type names

---

We introduce a notion of equivalence of type names and write T1 $\approx$ T2 to express that T1 and T2 represent the same carrier sets, i.e., CAR(T1) = CAR(T2). [4.6]

We moreover assume the typical operations on values associated with basic type names such as, e.g., logical connectives or arithmetic operators.

A special type name is Void, whose carrier set is unitary. [4.6]

---

**Definition**

- Void $\in$ UTYPE

- CAR(Void) = {void} with void $\in$ UVAL

---

*Variation Point:* Further basic type names – e.g., Real, Character or String and their subtyping relation, if any – are neither assumed nor detailed.

Type name constructors are simply functions that take one or more type names as argument and build a new type name. In the following sections, a number of type name constructors are introduced.

## 2.3   References and Variables

A reference is either Nil or an identifier for one value in the carrier set of a given type name. Let T be an arbitrary type name. Then Ref T is a type name whose carrier set consists of an infinite set of references including the distinguished reference Nil.

Given any type name T, the carrier set of type name Ref T has a very limited set of operations. References basically allow for comparison (i.e., test for equality), and dereferencing, and provide the special reference Nil. Given a reference r $\in$ CAR(Ref T), its dereference deref(r) $\in$ CAR(T) is defined iff r $\neq$ Nil.

---

**Definition** of references

- Ref: UTYPE $\rightarrow$ UTYPE

- Nil $\in$ CAR(Ref T) $\subseteq$ UVAL for any T $\in$ UTYPE

- deref: CAR(Ref T) $\rightarrow$ CAR(T) for any T $\in$ UTYPE
  with dom(deref) = CAR(Ref T) \ {Nil}

- deref: UTYPE $\rightarrow$ UTYPE
  with deref(Ref T) = T

**Notation**

- *r = deref( r )

---

**Remark:** The existence of the mathematically defined function deref from references to values does have an interesting effect. As mathematical functions do not change over time, references in Ref T for any type name T always refer to the same value. For this reason, we introduce below the concept of locations whose contents can change.

**Variation Point:** Comparison of references can be extended to include a "less than" relation, and even operations on references can be added in order to get a pointer arithmetic. For our purposes, however, these further relations and operations on references are not necessary.

In order to model records, objects, parameters and local variables of method calls and of executions, we introduce a notion of variable names, e.g., used to model attribute names in objects and names of local variables and parameters in methods. [4.7]

---

**Definition**

- UVAR is the universe of variables names (also called attributes)

- Each pair (T, u) with T $\in$ UTYPE and u $\in$ UVAR the denotes a typed variable of type T

- We denote typed variables also by u : T

---

## 2.4 Record Type Names and Cartesian Products

Type names can be composed into record type names. In addition to constructing new values from given ones, record values (i.e., values in the carrier set of a record type name) also provide selection functions for each part of a record value. In name-tagged records, these tag names provide proper names for the selection functions. [4.8]

---

**Definition** of records

- Rec: $\wp_f(\text{UVAR} \times \text{UTYPE}) \to \text{UTYPE}$
  where the variable names are all different

**Notation**

- $\text{Rec}(\{(a_1,T_1),(a_2,T_2),\ldots,(a_n,T_n)\})$ is denoted by $\text{Rec } \{a_1:T_1,a_2:T_2,\ldots,a_n:T_n\}$

**Definition**

- $\text{CAR}(\text{Rec } \{a_1 : T_1, ..., a_n : T_n\}) =$
  $\{ r : \text{UVAR} \to \text{UVAL} \mid r(a_i) \in \text{CAR}(T_i), 1 \le i \le n \}$

- attr: $\text{UTYPE} \to \text{UVAR}$  [4.8]
  $\text{attr}(\text{Rec } \{a_1 : T_1, ..., a_n : T_n\}) = \{a_1, ..., a_n\}$
  $\text{attr}(\text{Ref } T) = \text{attr}(T)$ for a reference type Ref T
  $\text{attr}(T) = \{\}$ for any other type

- proj: $\text{UVAR} \to \text{UVAL} \to \text{UVAL}$ with
  $\text{proj}(a_k)(r) = r(a_k)$       if $r \in \text{CAR}(\text{Rec } \{a_1 : T_1, ..., a_n : T_n\})$ and $1 \le k \le n$
  $\text{proj}(a_k)(r) = \text{proj}(a_k) (*r)$   if $r \in \text{CAR}(\text{Ref } T)$ and $a_k \in \text{attr}(T)$

**Notation**

- $\text{proj}(a_k)(r)$ is also denoted by $r.a_k$  if $r \in \text{CAR}(T)$ and $a_k \in \text{attr}(T)$
  $\text{proj}(a_k)(r)$ is also denoted by $r\text{->}a_k$ if $r \in \text{CAR}(\text{Ref } T)$ and $a_k \in \text{attr}(T)$

---

Any type names can be composed into record type names. The variables $a_i$ are called the attributes of the record type name. Notice that, as Rec is defined on (finite) sets of pairs, the definition of Rec does not rely on the ordering of its attributes, thus Rec {a:T, b:S} and Rec {b:S, a:T} are the *same* type name. If S and T are distinct type names, then so are Rec {a:S} and Rec {a:T}. If, however, S and T have overlapping carrier sets, then Rec {a:S} and Rec {a:T} have common record values in their carrier sets.

Note that, although we do not explicitly forbid that a type name is both of form Ref T and Rec S, it can often implicitly inferred. E.g., if a≠b, then from attr(Rec{a:Int}) = {a} and attr(Rec{b:Int})={b} we infer that both record type names are not equal. Furthermore, records [a=3] and [b=3] are also distinct values.

Cartesian products (also called "cross products") over values and the carrier sets of the record type names introduced above do share some common structure. However, they also differ significantly enough, so that we do not identify them. Records have indexed value entries, whereas Cartesian products have an ordered list of values. Although in programming Cartesian products can be mimicked by records quite well, in our system model we need a Cartesian product for example to model parameters of messages and methods.

In the following, we use List(.), Stack(.) and Queue(.), which are just mathematical constructs and not type extensions for the UML. We allow ourselves to build finite lists over arbitrary mathematical elements, using [1, …, n] for lists, len(.) for the length of a list, list(index) for item selection, and other common operators.

---

**Definition** of Cartesian products

- Prod: List(UTYPE) $\rightarrow$ UTYPE

- CAR(Prod$\{T_1, ..., T_n\}$) = $\times_{(1 \leq k \leq n)}$ CAR($T_k$)

- The empty list leads to a special unit type "Prod{}" whose carrier has one single value "()"

**Notation**

- We use Prod$\{T_1, ..., T_n\}$ as a synonym for the type and write:

  - $(p_1,...,p_n) \in$ CAR(Prod$\{T_1,...,T_n\}$) for a tuple of values ($p_i \in$ CAR($T_i$))

- There are two mappings between tuples and corresponding records:

  - rec$[a_1,...a_n]$ $(v_1,...v_n)$ = $[a_1=v_1,...a_n=v_n]$

  - prod$[a_1,...a_n]$ $([a_1=v_1,...a_n=v_n])$ = $(v_1,...v_n)$

---

The mappings between tuples and records are inverse. Note that both, tuples and records, need the list of attribute names: rec needs the ordered list $[a_1,...a_n]$ to map the values to the appropriate attributes, and prod needs the list to restore the order given in the tuple (which is not present in the record).

## 2.5 Locations

Static and dynamic parts are to be kept strictly apart. We thus introduce an explicit concept of locations for mutable values. The value stored at the location depends on the state of the state machine (see below). A location is an abstract representation of a part of the system store. [4.9]

---

**Definition** of locations

- ULOC $\subseteq$ UVAL is the universe of locations

- Loc: UTYPE $\rightarrow$ UTYPE
  Loc T denotes the type of locations that store data of type T

- CAR(Loc T) $\subseteq$ ULOC

---

For any type name T, we denote by Loc T the type name whose associated values are locations for values associated with type name T. Note that we allow arbitrary combinations of types such as Loc Ref T or Ref Loc T.

By ULOC $\subseteq$ UVAL we allow locations to be passed around and stored like ordinary values. The dereferencing of the location to the contained value is done in the context of the state represented by a store as defined below.

Note that by the explicit introduction of locations, the difference to references is made very clear. A reference points to a value and this relation is static, i.e., independent of the state of the system. A location contains a value (or its content) and is dependent on the state.

The function rvalue that retrieves the information stored in a location, as opposed to deref, is state dependant and thus its definition is delayed until the dynamic part of the state machines of the system model has been introduced. [4.9]

## 2.6 Class Names and Objects

Given a number of more traditional mathematical prerequisites, we now build the notion of objects and classes on top, basically by extending the operations available on certain type names and by restricting the way the elements associated to certain type names are structured and be used.

A class name defines attributes and methods, and may be related (by relationships) to other class names. Let C be a class name. We do not currently consider methods defined in class names, i.e., we consider them gathered in method suites and deal with them later.

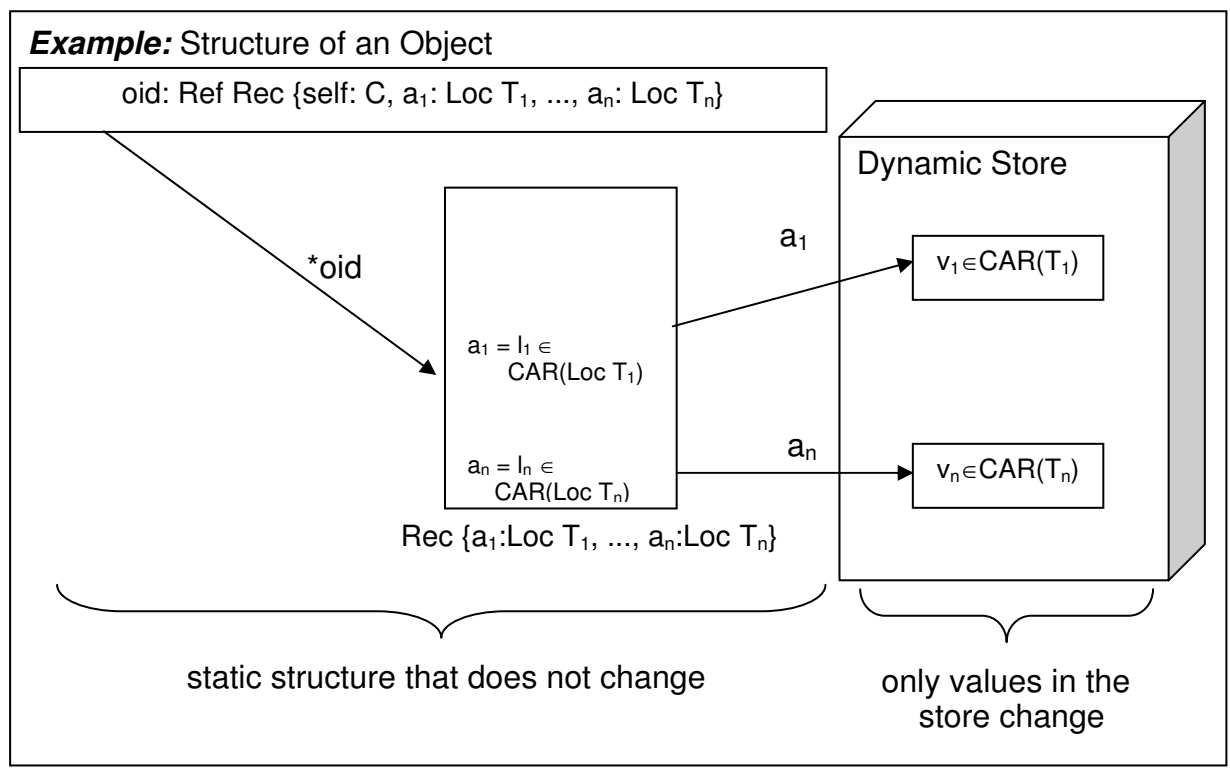

**Example:** Structure of an Object

oid: Ref Rec {self: C, $a_1$: Loc $T_1$, ..., $a_n$: Loc $T_n$}

**Fig. 3** Typical structure of an object

As Fig. 3 shows, we have a two stage dereferencing from the object identifier to the actually mutable attribute values. The object identifier at first references the static structure, which contains a number of locations for the attribute values in the data store.

Class names have a special structure associated. Each class denotes a set of object identifiers, characterised by a type name of the form:

$$C = \text{Ref Rec } \{self: C, a_1: \text{Loc } T_1, ..., a_n: \text{Loc } T_n\},$$

and their associated values, which are called objects, are associated to a type name of the form:

$$*C = \text{Rec } \{self: C, a_1: \text{Loc } T_1, ..., a_n: \text{Loc } T_n\}$$

Thus a class name is a type name as well.

CAR(C) denotes the set of object identifiers for the class C. A single dereferencing of those identifiers leads to object structures:

$$\text{CAR}(C) \subseteq \text{UOID}$$

$$\text{CAR}(*C) = \{ *oid \mid oid \in \text{CAR}(C) \} \subseteq \text{INSTANCE}$$

(UOID and INSTANCE are formally defined below.) Object identifiers uniquely point to object structures (formally records) and we do not have dangling references, so there is a bijection between object identifiers and object structures, i.e., dereferencing (*.) is bijective (when disregarding Nil).

Finally, we do not allow objects to share locations[6]. So for any two object identifiers $o_1 \in \text{CAR}(C_1)$, $o_2 \in \text{CAR}(C_2)$ and attribute names $a \in \text{attr}(*C_1)$, $b \in \text{attr}(*C_2)$ we have $*o_1.a \neq *o_2.b$ for any a and b with Loc-types.

Note that UOID contains references to all possible objects and, in a similar way, INSTANCE contains all possible objects in a data store.

Through the dereferencing function object identifiers know about the objects they dereference. However, objects also have some knowledge about themselves. This means, an object knows its identifier and its class. As a consequence of this definition each object belongs exactly to one class.

---

[6] Reusing object locations of terminated objects leads to a sharing of locations in implementations for pure storage optimization and need not be modelled in the conceptually abstract system model.

**Definition** of classes and instances:

- Ref Rec {self: C, $c_1 : T_1$, ..., $c_k : T_k$, $a_{k+1}$: Loc $T_{k+1}$, ..., $a_n$ : Loc $T_n$}, is a class name

- UCLASS $\subseteq$ UTYPE is the universe of class names,

- UOID = $\cup_{C \in UCLASS}$ CAR(C)  is the universe of object identifiers,

- INSTANCE = $\cup_{C \in UCLASS}$ CAR(*C)  is the set of objects,

  where for each C $\in$ UCLASS there are unique $a_i$, $T_i$ such that
  C = Ref Rec {self: C, $a_1$: Loc $T_1$, ..., $a_n$:Loc $T_n$},

  and for all oid $\in$ UOID the object identifier fits:
  *oid.self = oid

- classOf: INSTANCE $\rightarrow$ UCLASS

  with object $\in$ CAR(classOf(object))

- classOf: UOID $\rightarrow$ UCLASS

  with classOf(oid) = classOf(*oid)

Fig. 4 illustrates our semantic universe.

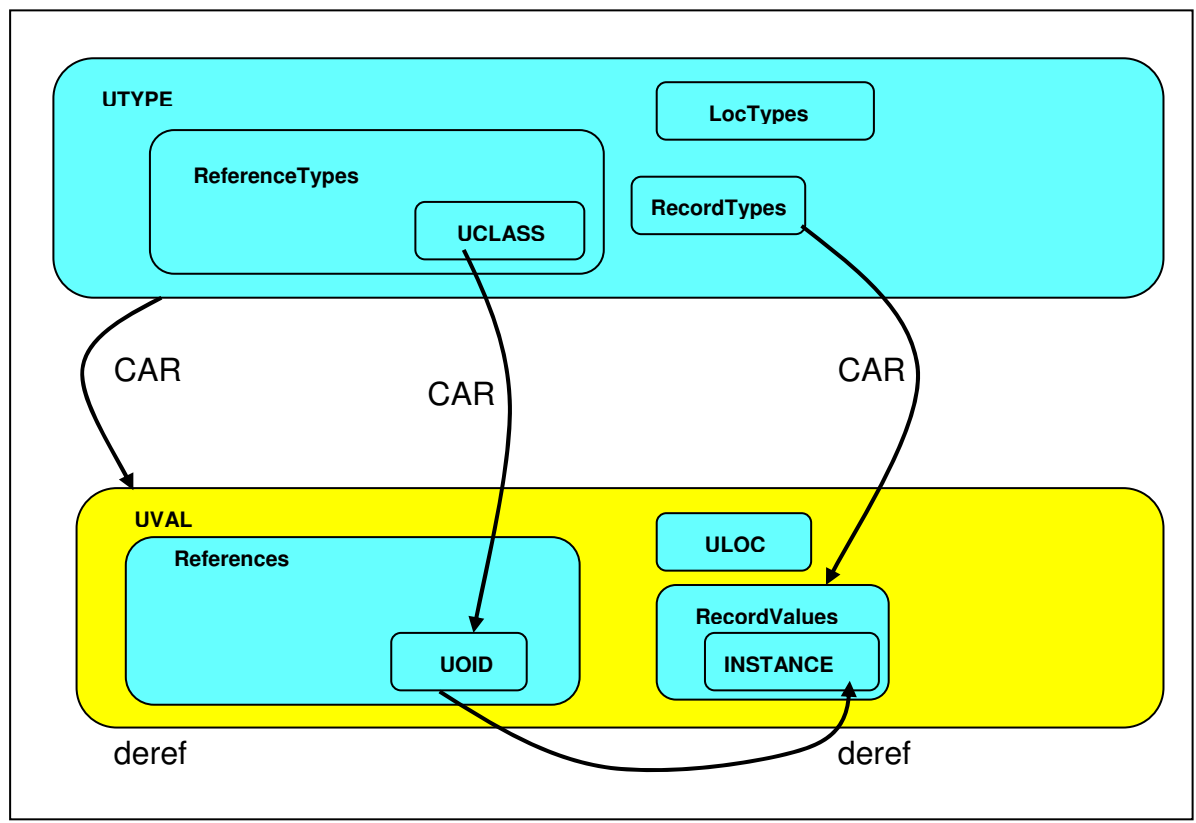

**Fig. 4** Overall picture of the semantic universe

## 2.7 Subclassing

*Subclassing* (also called *Inheritance*) is a basic feature in object-oriented programming. To indicate that a class C' inherits from a class C, we introduce binary subclass relation "sub" on the universe of types  which implies a number of conditions on related classes:

---

**Definition** of subclassing:

- sub $\subseteq$ UCLASS $\times$ UCLASS is the transitive and reflexive "subclass" relation

- Oid: UCLASS $\rightarrow$ UTYPE is a type constructor for the type of object identifiers for class C including its subclasses

  defined by CAR(Oid(C)) $= \cup_{\text{C1 sub C}}$ CAR(C1) $\subseteq$ UOID

---

The above definition is sufficient to capture subclassing on the structural side. However, it also leaves quite a few things open. For instance the binary relation sub is not enforced to be antisymmetric[7] (although no implementation language supports this today).

From the definition, we can see that in UML a subclass can have an arbitrarily different record structure. Furthermore, subclassing is not based on a structural definition: two classes C1 and C2 may have the same attributes, but still be in no relationship at all.

The substitution principle [LW94] enforces object identifiers of subclass C1 to be special cases of class C. That can easily be modelled by introducing a set inclusion on object identifiers, since we do not pass around objects, but identifiers we do not need the set inclusion on objects, but on the identifiers.

With this technique on defining a subset relation on object identifiers instead of objects, we get a great simplification on the mimicked typing system. Furthermore, it allows us to redefine attribute structures in subclasses without a loss of the substitution principle.

**Remark:** *Multiple inheritance* allows a class to inherit features from more than one class. While a constructive class definition inherits from several classes, a relational point of view just resolves that as several binary inheritance relationships and is therefore already covered in our definition.

To avoid name conflicts that arise when attributes of different superclasses (or of a superclass and of the extension) are homonyms, we require that all attribute definitions introduce different names. Semantically, this convention means no restriction, because attribute access is always resolved statically at runtime and there is no dynamic lookup on attributes.

---

[7] A binary R $\subseteq$ S $\times$ S is called antisymmetric if $\forall s_1, s_2 \in S$: $s_1$ R $s_2 \wedge s_2$ R $s_1 \Rightarrow s_1 = s_2$.  A classical example for such an R is the "less than or equal to" relation on natural numbers.

***Variation point:*** In procedural languages, redefinition of attributes in subclasses is not possible without a loss of type safety. As a semantic variation, we may enforce type safety, by requiring subclasses to keep their attributes and these attributes to keep their names and therefore types. Therefore a subclass only extends the record structure of its superclass:

for any classes C1 sub C2 we have attr(C2) $\subseteq$ attr(C1)

***Extension point "dynamic reclassification":*** Notice that, an object may be regarded as instance of more than one class along the subtyping hierarchy. Thus an object may be dynamically reclassified by its context according to the given subclass hierarchy. However, this only changes the external viewpoint of an object, but neither its internal structure (existing attributes) nor its behaviour.

In UML 2.0 dynamic reclassification for classes is introduced in a very general way. The system model does not reflect this capability of reclassification, because we assume that this concept should be mapped to the system model through introduction of additional infrastructure. E.g. possible implementations of dynamic reclassification go along building an additional superclass that contains all attributes and a flag which behaviour is currently active. Even more flexibility becomes possible, when chance of dynamic behaviour is realised through delegation of behaviour to other objects.

## 2.8  Extension and Variation Point: Type system

Further constructs for building type names are possible. For instance, an array type name or a subtyping structure beyond the subclassing concept inherent in OO may be available. We also did not deal with parametric polymorphism, for example, which was introduced in Java 1.5 in form of instantiable templates within the system model. A type system is an enhanced syntactic concept and should therefore be handled together with the concrete syntax of the models.

The store model that we introduce does not provide and, as it is a purely semantic construct, does not need any explicit visibility or hiding mechanisms. In particular, it does not describe which activities may change which attributes. Although it is recommended from engineering practice to prevent foreign objects from changing an attribute (and thus use private attributes only), passing references can be modelled. This allows in/out-parameters, but enables and even encourages capsule leaks.

Any attribute attr with type Loc Loc T contains a location for a location. Locations are ordinary values and can be stored, passed around, as well as used for reading and writing. Furthermore, the located place of value T in the store can be changed as well.

## 2.9 The Data Store Structure

In the system model, we abstract away a number of details, such as storage layout and physical distribution. We use an abstract global store to denote the state of an object system. Even if there is no such concept in the real (distributed) system, all instances are organised in this single global store.

Intituively, the data store models the state of a system at a certain point in time. At each point of time the store contains real objects for a finite subset of the universe UOID of all object identifiers. Time progress is modelled by state transitions of the state machine mentioned in Sect. 1.4, which is defined in a separate document.

Please note that even though we have conceptually defined a global data store, we do not enforce the existence of such a globally defined state in the implementation. We also allow interleaving, as well as concurrent activities, as can be seen in the next behavioural part of the system model. This will be detailed further when we define the dynamic parts, such as control store, stack and processes.

A data store is a snapshot of the data state of a running system. Stores contain assignments to locations and describe the currently instantiated set of objects.

---

**Definition** of the data store:

In the system model, let

- DataStore = $\wp$(UOID) $\times$ (ULOC $\rightarrow$ UVAL)  the set of snapshot values
- oids: DataStore $\rightarrow \wp$(UOID) the set of existing objects

    where oids((s,m)) = s
- locations: DataStore $\rightarrow$ ULOC the set of used locations

    where locations((s,m)) = dom(m)

---

There are quite a number of restrictions on DataStore:

The elements of DataStore contain partial mappings since only a finite number of locations may actually be in use in any snapshot of the computation. Furthermore, the locations used in the existing objects correspond to the locations used in the store at that moment. Therefore both the set oids(store), which contains the existing object identifiers of the DataStore store, and the set locations(store), which contains the existing locations of store, are finite.

It is necessary to have a number of retrieval and update functions for the data store at hand. Those are defined below:

**Definition** DataStore functions:

For the data store, let these retrieval functions be defined:

- val: DataStore $\times$ ULOC $\rightarrow$ UVAL retrieving the value for a given location
  val((s,m), loc) = m(loc)

- val: DataStore $\times$ UOID $\times$ UVAR $\rightarrow$ UVAL retrieving the value for a given object and attribute
  val((s,m), oid, at) = m(*oid.at)

- vals: DataStore $\times$ UOID $\rightarrow$ (UVAR $\rightarrow$ UVAL) retrieving the mapping of attribute name to value for a given object
  vals((s,m), oid) = { f | dom(f) = attr(classOf(oid)) $\wedge$
  $\qquad\qquad\qquad\qquad\qquad\qquad$ $\forall$at$\in$dom(f): f(at) = m(*oid.at) }

Furthermore, the following updates can be used to define changes:

- addobj: DataStore $\times$ OID $\times$ (ULOC $\rightarrow$ UVAL) $\rightarrow$ DataStore   adding an new object
  addobj((s,m), oid, f) = (s$\cup${oid}, m$\oplus$f)

- setval: DataStore $\times$ ULOC $\times$ UVAL $\rightarrow$ DataStore   setting the value for a location
  setval((s,m), loc, v) = (s, m$\oplus$[loc=v])

- setval: DataStore $\times$ UOID $\times$ UVAR $\times$ UVAL $\rightarrow$ DataStore  setting the value for a given object and attribute
  setval((s,m), oid, at, v) = (s, m$\oplus$[*oid.at=v])

As shorthands, we use

| | | |
|---|---|---|
| ds(loc) | for | val(ds,loc) |
| ds[loc = v] | for | setval(ds,loc,v) |
| ds(oid.at) | for | val(ds,oid,at) |
| ds[oid.at = val] | for | setval(oid,at,val) |

Again various restrictions on the use of retrieval and update functions apply. This involves the use of values of appropriate type, attributes that actually exist in a class, etc. Furthermore, a number of properties can be derived e.g., from selected values from the typing information. However, we refrain from defining these restrictions at the moment.

At each point in time, i.e., in each state of the state machine, when the instance exists, we assume that its attributes are present and the values in these locations do have defined values (including Nil), but it is not necessarily the case that we do know about these values. They may be left underspecified. In particular it may be that, after creation of an instance, its attributes still need to be initialised. Please note that is a nice modelling technique used e.g., in verification systems to avoid to explicitly handle a pseudo-value "undefined" [NPW02]. It also resembles reality, e.g., when there is an un-initialized variable of type "int". When accessing the value, we do know that it contains an integer, but we do not have any clue which one it is.

On the assumption that we can use a consistent global state for all instances at each point of time, even if they are in computational activities, we can model the global data state of an object system by this data store.

### 2.10 Class Variables and Constants

While attributes are by far the most commonly used elements to store values, there are three further types of elements present in the object-oriented universe.

Constants on one hand are values with a name, such that the name can be used instead of the value. We do not need to represent constants explicitly in the system model: Their associated values are present in the universe of values and the mapping of names to values as well as their visibility is not part of the system model, but part of the mapping from UML to the system model.

A second concept that we have not explicitly represented so far is the concept of static attributes. These are attributes that can be regarded as shared between all objects of the class. Indeed they exist independently of any object, but can only be accessed from within a limited scope. While the system model does not cope with visibility of a static attribute, it is prepared to incorporate the static attribute by assigning a location to it that is not part of any object. This way the system model is able to deal with static attributes.

Please also note that a different, but also convenient way to include a static attribute into all objects of a class is to include its location in all objects uniformly, thus allowing objects to share the location.

### 2.11 Associations: Various Object Relations

One of the core elements of UML is the concept of an association. Associations are relations between object identifiers. While most of them are binary, associations may be of any arity, may be qualified in various ways and may have additional attributes on their own. Furthermore, associations can be "owned" by one or more of the participating objects/classes or it me stand on it's own, not owned by any of the related objects. In implementations a basic mechanism for managing those relations is to use direct links or Collection classes, but there are other possibilities as well. To allow for different variants of realisations of associations, we use a generalised, extensible approach: We use relationship identifiers to extract links from the store and allow for a variety of realisations of these functions. This approach is very flexible, as it, on the one hand, abstracts away from the owner of associations as well as the form how associations are stored and, on the other hand, does not restrict any possible form of an association. As a disadvantage of this approach, we cannot capture all forms of associations in one uniform characterisation, but do provide a number of standard patterns that cover the most important cases. If no standard case applies, e.g. for a new stereotype of an association, then the stereotype developer has to describe his interpretation of the stereotype directly in the terms of the system model given below. For simplicity, we demonstrate this approach by defining variants of binary associations below.

In general any association has a name R, a signature given by a list of classes $(C_1,\ldots,C_n)$, possibly additional attributes of that association and a relation retrieval function relOf(R): DataStore $\rightarrow \wp_f (CAR(Oid\ C_1) \times \ldots \times CAR(Oid\ C_n) \times UVAL^k)$.

Note that the use of $CAR(\text{Oid } C_1)$ includes relations between objects of subclasses of $C_i$, which is usually intended by associations, but not covered if we would have used carrier sets $CAR(C_i)$ of objects belonging directly to $C_i$.

Please also note that with this approach it is possible to model qualified associations by interpreting one (or more) of the additional attributes as the qualifier as well as to model non-unique associations by introducing a value as distinguishing flag. Some example for association mappings are given below by starting with a binary association.

As a third, you should also note that associations usually define certain restrictions on their changeability. This cannot be stated within the state part of the system model, but only when sequences of DataStores are used to compare behaviour over time.

As said, the retrieval function relOf is depends on the concrete realisation of the association. Even after quite a number of years of studying OO formalisations, there is so far not really a satisfying approach describing all variants of association implementations. Therefore, we provide this abstract function and impose certain properties on the function, without discussing the internal storage structure. The only decision we made so far is that associations are somehow contained within the store, i.e., they are somehow part of objects and locations and association relations do not extend the store. This is pretty much in the spirit of the system model, where higher-level concepts are explained using lower level concepts.

---

**Definition** of associations:

In the system model, let

- UASSOC be the universe of association names and

- relOf(R): UASSOC $\rightarrow$ DataStore $\rightarrow$
  $$\wp_f(CAR(\text{Oid } C_1) \times \ldots \times CAR(\text{Oid } C_n) \times UVAL^k)$$
  the retrieval function to derive the actual links for an n-ary association based on the current store.

---

Please note that we do not constrain the relationship between UASSOC and UTYPE. In particular one might regard each association manifesting itself as type (UASSOC$\subseteq$UTYPE) or only a few associations being realised as types. This for example allows us to model simple relations as attributes only, without having to attach a type to them within the system model as shown below.

***Extension point:*** Many associations are binary without any additional attributes. For those we can use

$$\text{binaryRelOf: UASSOC} \rightarrow \text{DataStore} \rightarrow \wp_f(CAR(\text{Oid } A) \times CAR(\text{Oid } B))$$

as a retrieval function to derive the actual links.

**Extension point:** The above given retrieval does not yet regard ordering, which is necessary for handling associations with the ordering stereotype, nor qualified associations, nor does it handle the possibility that a pair of objects is linked several times (multiple associations). An appropriate extension for ordering is given by the more detailed retrieval function

$$\text{orderedBinaryRelOf: UASSOC} \rightarrow \text{DataStore} \rightarrow$$
$$\text{CAR(Oid A)} \rightarrow \text{CAR(List(Oid B))}$$

Similarily, for a qualified relation we define:

$$\text{qualifiedBinaryRelOf: UASSOC} \rightarrow \text{DataStore} \rightarrow$$
$$\wp_f\,(\text{CAR(Oid A)} \times \text{CAR(Oid B)} \times \text{CAR(Q)}),$$

where type name Q is the qualifier that identifies unique B-objects starting in an A-object.

**Variation and extension point:** Retrieval functions are not further specified yet, because they may have quite a number of different realisations. For clarification, we define a few below, covering standard cases and provide them as variation points. However, more variations are possible, such as allowing us to introduce our own variants. Therefore, this is also an extension point.

---

**Defining** a simple binary relation "SimpR":

The simplest kind of association "SimpR" from A to B is realised unidirectionally in class A through an attribute "simpR" to link to the other side:

- The structure of *A looks like Rec(…, simpR: Loc B, ... ) and

- the retrieval function is defined by:

    binaryRelOf(SimpR) (ds) =
    $\{\,(x,y) \in (\text{CAR(Oid A)} \times \text{CAR(Oid B)}) \mid y = ds(x.simpR)\,\}$

To illustrate that, one might think of a transformation of the following kind, to realise SimpR:

---

The following definition for a binary *-to-*-association works quite similarly for n-ary associations with arbitrary n:

**Defining** a *-to-*-binary relation "Med":

*-to-*-association "Med" uses an intermediate class also called "Med" to store its links. It does not include the association state in any of the objects of A or B:

- *Med looks like Rec(…, a: Loc A, b: Loc B, ...) and the

- retrieval function is defined by:

    binaryRelOf(Med) (ds) = { (x,y)∈ (CAR(Oid A) × CAR(Oid B)) |
                ∃ m∈ CAR(Med):  x = ds(m.a) ∧ y = ds(m.b)  }

To illustrate this, one might think of a transformation of the following kind, to realise Med:

Model:
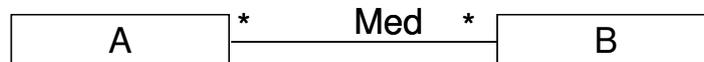

Realized through:
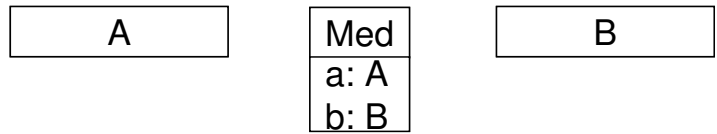

The above two definitions demonstrate that the issue of owning a link can quite generally be covered through the use of abstraction functions. In the first definition, the objects own the links, in the second, the links are separated from the objects. Of course also combinations are possible, as shown in the following third definition.

---

**Defining** a redundant binary relation "Med":

The *-to-1-association "Med" has redundant links. Participating objects from A, directly point to B objects. From B the association is realized using an external collection:

- let us assume class Collection(A) provides a function
  vals: DataStore × CAR(Collection(A)) → $\wp_f$( CAR(Oid A) )

- then our retrieval function is simply defined by:

  binaryRelOf(Med) (ds) =
  { (x,y)∈ (CAR(Oid A) × CAR(Oid B)) | y = ds(x.med) }

  or as an equivalent alternative by:

  binaryRelOf(Med) (ds) =
  { (x,y)∈ (CAR(Oid A) × CAR(Oid B)) | x ∈ vals(ds,y.med) }

As the realization is a redundant data structure, we enforce the consistency constraint that both sets are equal. To illustrate this situation, one might think of a transformation of the following kind, to realise Med:

Model:

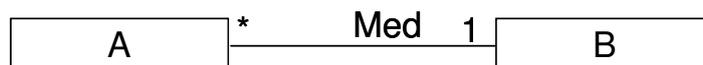

Realized through:

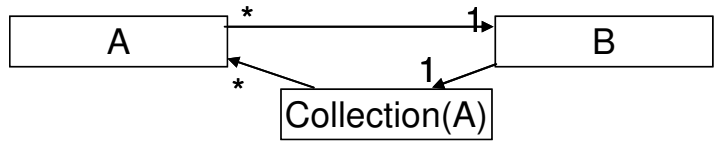

---

***Variation and extension point:*** If collections are used in an implementation the corresponding retrieval functions are an abstraction of what collections actually store. Note that these functions are mathematical constructs that make the intentions of collection classes explicit, but need not be actually implemented.

It is important to note that the effect e.g. of an action on the links of an association can be described by using the retrieval function, without having to actually look at the actual representation in the system model. It is not even necessary to provide such a representation, but suffices to know there is one. This principle comes from abstract types in algebra, where the changes on data structures are also purely defined on the effect on access functions.

# 3   References

[AHU83]   Alfred V. Aho, John E. Hopcroft, and Jeffrey D. Ullman. *Data Structures and Algorithms*. Addison Wesley Publishing Company, 1983.

[BG92]   Gérard Berry and Georges Gonthier. The ESTEREL Synchronous Programming Language: Design, Semantics, Implementation. *Science of Computer Programming*, 19(2):87—152. Elsevier North-Holland, 1992.

[BKR96]   M. Broy, C. Klein, B. Rumpe. A stream-based mathematical model for distributed information processing systems – SysLab system model –. In: Proceedings of the first International Workshop on Formal Methods for Open Object-based Distributed Systems. Chapman-Hall. 1996.

[BS01]   Manfred Broy and Ketil Stølen. *Specification and Development of Interactive Systems: Focus on Streams, Interfaces, and Refinement.* Springer, 2001.

[EHS97]   Jan Ellsberger, Dieter Hogrefe, and Amardeo Sarma. *SDL: Formal Object-Oriented Language for Communication Systems.* Prentice Hall, 1997.

[GR95]   Radu Grosu and Bernhard Rumpe. *Concurrent Timed Port Automata.* Technical Report TUM-I9533, Technische Universität München, 1995.

[HR04]   D. Harel, B. Rumpe. Meaningful Modeling: What's the Semantics of "Semantics"? In: Computer, Volume 37, No. 10, pp 64-72. IEEE, October 2004.

[Hoa78]   C.A.R. Hoare. Communicating Sequential Processes. Communications of the ACM, 21(8):666—677. ACM Press, 1978.

[Kah74]   Gilles Kahn. The Semantics of a simple language for Parallel Programming. In *IFIP Congress'74 (Proceedings)*, pages 471—475. North Holland Publishing, 1974.

[Kat93]   Randy H. Katz. *Contemporary Logic Design*. Addison Wesley Publishing Company, 1993.

[LP99]   J. Lilius and I. Porres. Formalising UML state machines for model checking. In R. B. France and B. Rumpe, editors, The Unified Modeling Language (UML 1999), volume 1723 of Lecture Notes in Computer Science, pages 430-445. Springer, Berlin Heidelberg New York, 1999.

[LW94]   Liskov and J. Wing. Family Values: A Behavioral Notion of Subtyping. ACM Transactions on Programming Languages and Systems, November 1994.

[Mil80]   Robin Milner. A Calculus of Communicating Systems. In vol. 92 of *Lecture Notes in Computer Science.* Springer, 1980.

[NPW02]   Tobias Nipkow, Lawrence C. Paulson, Markus Wenzel. Isabelle/HOL. Springer, Berlin. 2002.

[OMG04] Object Management Group. Unified Modeling Language: Superstructure, version 2.0, formal/05-07-04.

[Win93]   Glynn Winskel. *The Formal Semantics of Programming Languages: An Introduction*. MIT Press, 1993.

# 4   Appendix: Rationale and Comments

## 4.1  General Approach to Semantics

The semantic mapping is of the form:

Sem: UML  $\rightarrow$   $\wp$ (Systemmodel)

and functionally relates any item in the syntactic domain to a construct of the semantic domain.  The semantics of a model m $\in$ UML is therefore Sem(m).  Due to compositionality, given any two models m,n$\in$UML combined into a complex one $m_{\oplus}n$ (for any composition operator $\oplus$ of the syntactic domain), the semantics of $m_{\oplus}n$ is defined by Sem($m_{\oplus}n$) = Sem(m) $\cap$ Sem(n). In the same way, n$\in$UML is a (structural or behavioural) refinement of m$\in$UML, exactly if Sem(n)$\subseteq$Sem(m). Formally, refinement is the nothing else than "n is providing at least the information about the system that m does". These general mechanisms provide a great advantage, as they simplify any reasoning about composition and refinement operators.

## 4.2  Structuring the Semantics of UML

UML provides a number of derived operators which do not enhance the expressivity of the language but the comfort of its use. Derived constructs can be defined in terms of constructs of the core as, e.g., state hierarchy of UML's state transition diagrams can be neglected without losing expressivity.

The onion approach described in Sect. 1.2 can be used in two ways: It is undoubtedly good if and when the UML sublanguages can be organised into a hierarchy – one sublanguage is more concrete than another one.  However, the UML (sub)languages mostly present different views on the same whole, thus making the onion metaphor less adequate. Nevertheless, transformations within a sublanguage that replace derived constructs, e.g., flattening hierarchical states of a state transition diagram, make the task easier since we have to deal only with a subset of the language.

The various views on a system that can be offered by the different UML sublanguages are all mapped onto predicates on the system model. That is, the system model supports both view integration and model consistency verification.

## 4.3  The Math behind the System Model

Actually, there is a crucial difference between the precise properties of the elements in the system model and their representation. Different representations of the same system model are possible, of course, but there is a necessity to define syntax, semantic domain, and semantic mapping in a precise way. To concentrate on the essentials, we use the abstract syntax of the UML. Whereas the UML itself is graphical in nature, we do not deal with pixels forming boxes. Abstract syntax can, for example, be described using "metamodelling", but for a variety of reasons it is not suitable for us. In particular, it is neither elegant to deal with when defining semantics because we need to build an infrastructure above meta-models anyway, nor is it as formal as necessary, which would force us to formalize meta-modeling techniques first. We instead rely on a mathematical version. In the literature, Z, B, CSP, or other already existing formal languages are used, usually permitting a good description of a specific part of UML while unable to describe other parts. Mathematics however, is the most powerful tool, flexible and general, and actually needed in order to give semantics to UML 2.0.

We also put some emphasis on advantageous semantic and algebraic properties of the system model. As an example, in the system model, interface abstraction distributes over system composition and is thus called a morphism. This might give rise to mirroring this property in UML, thus enhancing the UML 2.0.

## 4.4 Static and Dynamic Issues

An object-oriented system can basically be described using one of various existing paradigms. One paradigm is that of a set of communicating state machines, another that of a single, called global, state machine. In this latter case, the global state machine, if detailed enough, is perfectly appropriate to model parallel, independent and distributed computations.

In principle, a system of communicating, elementary state machines could be considered more convenient than a single, global machine for describing the semantics of UML models. It is also possible to construct a global state machine by integrating elementary ones; however, this is a non-trivial operation. Therefore, it is more appropriate to employ the concept/metaphor of one state machine at a higher, non-elementary level. In fact, we introduce a composition operator on state machines representing fragments of larger systems, such that these state machines can be composed, leading to larger state machines.

The static information contained in the static part of the global state machine includes the definition of classes including (the definition of) their methods; this is not explicitly treated in this report. Also, references are defined as static.

The dynamic information contained in the dynamic part (i.e., in the states) of the global state machine includes references and the values of their locations.

The control structure contained in the control part of the global state machine may include, besides the invoked methods, also signals and actions emitted by UML's state transition diagrams or interactions, dispatched and not yet delivered messages passed from one object to another one, actions of an activity diagram, etc. These issues will be dealt with when defining the semantic mapping for each one of the languages of UML.

The state transition function may record some useful information, for instance, when a state transition took place and what triggered the transition.

In the database realm, the static part is called "schema", and the dynamic part is the "instance". The schema instantiation is changeable while the schema itself is not. Schema changes (usually called "schema evolution" in the literature) are not considered, as they usually do not occur within a running system, but when evolving and/or reconfiguring it.

## 4.5  Types

The word "type" simultaneously has two meanings. On the one hand, within a UML model, a type is a name lacking a formal semantics and intuitively understood as a type of any (object-oriented) programming language, whose members do not own an identity, and which is characterised by the operations it has associated. On the other hand, within the system model we also have a notion of type used to conveniently describe sets of various kinds (far beyond the UML notion of type).

Although we do not deal with peculiarities of various type systems, strong or weak typing, etc., we outline the underlying type system, as we need to map the type information of UML to this type system.

The universe UTYPE of type names in the system model is not detailed further. Although $T \in$ UTYPE models a type, T actually stands for a name, and in short we say type T for it. In that respect, we use a deep embedding of the type system of UML, by representing it through type names and a universe of values only. By deep embedding, we mean that we do not map types of the UML to a type system of the underlying mathematical structure, but explicitly model types as first-class elements.

Carrier sets need not be disjoint. This notion of type allows the subsumption of object types and value types as well as reference types. We may however enforce values (or just members of certain types) to have a single (or most specific) type, for instance by means of a function typeOf: UVAL $\rightarrow$ UTYPE, a partial assignment of a type for each value. In ordinary object-oriented programming languages, objects usually have an assigned type (even though there is subtyping, the assigned type is the class the object is instance of), but special values like Nil usually do not.

A variant to a typeOf function, especially suited when no default type is to be assigned to values and the carrier sets are not disjoint, is the introduction of values paired with their type information (e,T) such that $e \in$ CAR(T). So for instance the number 3 can be regarded as an integer value in (3,Int) or as a real value in (3,Real).

## 4.6  Basic Types and Type Constructors

The value void is usually needed for giving semantics to procedures or methods with no return value. This is customary in the semantics of programming languages.

Type constructors, as we have introduced them in Sect. 2, allow us to build a language of types. For example, we can introduce a record constructor Rec{.,.} and state that any two record types are equivalent if they have the same constituents independent of their order: Rec{A,B} $\approx$ Rec{B,A}. This of course demands a certain discipline when defining CAR, so that two equivalent types indeed have the same carrier set associated.

There are a number of other approaches, for instance the constructor Rec(.,.) (note the use of parenthesis instead of curly braces) may be such that the ordering of its arguments do matter. Then, Rec(A,B) and Rec(B,A) are different (i.e., not equivalent) types, but we may state them to be isomorphic. This can mean that the associated carrier sets are not identical but related by a bijection. For our purposes, however, this concept of isomorphism is imprecise as long as we do not speak about algebras but only about type names.

## 4.7  References and Variables

Formally, deref can be defined as a type dependent function

$$deref: T{:}UTYPE \rightarrow CAR(Ref\ T) \rightarrow CAR(T)$$

or as function with an additional restriction

$$deref: UVAL \rightarrow UVAL$$
such that any $v \in CAR(Ref\ T)$, $v \neq Nil$, is mapped to $deref(v) \in CAR(T)$

or as a family of functions $deref = (deref_T)_{T \in UTYE}$ indexed by types with

$$deref_T:\ CAR(Ref\ T) \rightarrow CAR(T)$$

In ordinary programming languages, variables shadow each other when a new variable with the same name is introduced in an inner scope. We assume static binding, thus each variable name can be statically resolved (as opposed to dynamic binding, by which the resolution of a variable name depends not on the environment of its definition but on the environment of its use, and thus variable resolution can only occur at run time). In other words, in the modelling languages we deal with, we assume that a consistent, model-wide redefinition of variable names is possible in such a way that each variable is used only once. Then variable shadowing does not occur, and need not be dealt with in the system model.

## 4.8  Record types

Record types are a classical concept to represent the state space of classes and their objects. In particular, in name-tagged records, these functions do have appropriate names, resembling attribute names.

If values have types associated by the function typeOf (see Sect. 4.5), then the function returning the attributes of a record type can be extended to values as follows:

$$attr: UVAL \rightarrow \wp(UVAR)$$
$$attr(v) = attr(typeOf(v))$$

## 4.9  Locations

In implementation-oriented terms, locations are also often called *lvalues*. In each system state, a value (also often called *rvalue*, the content of the location at a given point in time) is associated with it. In an expression "x := x+1", the first occurrence of the variable x denotes its location while the second one its rvalue that is derived from the store by knowing the location to look at.

We could also work with an approach where we do not introduce variable identities explicitly. Such an approach would be slightly simpler, but less expressive. So far we do not use the extra expressivity explicitly, however. Another possibility would be to unify locations and references (using in our setting always a combined "Ref Loc" as pointer).